# Ballistic Ejection of Microdroplets from Overpacked Interfacial Assemblies[†]

Xuefei Wu[1,2], Gautam Bordia [2,3], Robert Streubel[4], Jaffar Hasnain[2,5], Cássio C.S. Pedroso[6], Bruce E. Cohen[6,7], Behzad Rad[6], Paul Ashby[6], Ahmad K. Omar[2,3*], Phillip L. Geissler[5†], Dong Wang[1*], Han Xue[8], Jianjun Wang[8], Thomas P. Russell[2,9,1,10*]

[1]Beijing Advanced Innovation Center for Soft Matter Science and Engineering & State Key Laboratory of Organic-Inorganic Composites, Beijing University of Chemical Technology, Beijing 100029, China

[2]Materials Sciences Division, Lawrence Berkeley National Laboratory, Berkeley, CA 94720, USA

[3]Department of Materials Science and Engineering, University of California, Berkeley, Berkeley, CA 94720, USA

[4]Department of Physics and Astronomy, University of Nebraska-Lincoln, Lincoln, NE 68588, USA

[5]Department of Chemistry, University of California, Berkeley, Berkeley, CA 94720, USA

[6]Molecular Foundry, Lawrence Berkeley National Laboratory, Berkeley, CA 94720, USA

[7]Division of Molecular Biophysics & Integrated Bioimaging, Lawrence Berkeley National Laboratory, Berkeley, CA 94720, USA

[8]Beijing National Laboratory for Molecular Science, Institute of Chemistry, Chinese Academy of Sciences, Beijing 100190, China

[9]Polymer Science and Engineering Department, University of Massachusetts, Amherst, MA 01003, USA

[10]Advanced Institute for Materials Research (AIMR), Tohoku University, 2-1-1 Katahira, Aoba, Sendai 980-8577, Japan

†deceased

[†]We dedicate this contribution to the memory of Phillip L. Geissler, an exceptional colleague, scientist and collaborator without whom this work would not have been possible.

*Corresponding author. Email: aomar@berkeley.edu; dwang@mail.buct.edu.cn; russell@mail.pse.umass.edu

**Abstract**

**Spontaneous emulsification, resulting from the assembly and accumulation of surfactants at liquid-liquid interfaces, is an interfacial instability where microdroplets are generated**

**and diffusively spread from the interface until complete emulsification. Here, we show that an external magnetic field can modulate the assembly of paramagnetic nanoparticle surfactants (NPSs) at liquid-liquid interfaces and trigger an oversaturation in the areal density of the NPSs at the interface, as evidenced by a marked reduction in the interfacial tension, γ, and corroborated with a magnetostatic continuum theory. Despite the significant reduction in γ, the presence of the magnetic field does not cause stable interfaces to become unstable. Upon rapid removal of the field, however, an explosive ejection of a plume of microdroplets from the surface occurs, a dynamical interfacial instability which is termed *explosive emulsification*. This explosive event rapidly reduces the areal density of the NPSs to its pre-field level, stabilizing the interface. The ability to externally suppress or trigger the explosive emulsification and controlled generation of tens of thousands of microdroplets, uncovers an efficient energy storage and release process, that has potential applications for controlled and directed delivery of chemicals and remotely controlled soft microrobots, taking advantage of the ferromagnetic nature of the microdroplets.**

**Main Text**

The use of paramagnetic nanoparticle surfactants (NPSs) provides additional control of surfactant adsorption through the application of an external magnetic field[1-5] (in addition to the more canonical control variables of surfactant concentration and solution chemistry e.g., pH[5-8]). Magnetic field gradients are widely used to steer the movement of objects[9] or to realize function, including rotation, swimming, rolling and transport of cargo[10]. A magnetic field also induces a magnetic dipole moment in superparamagnetic particles. When the interparticle dipole interaction energy is large enough to overcome thermal energy, the dipolar force drives the assembly of magnetic particles into chain-like structures at low particle concentrations[11-13], where the interparticle distance can be finely tuned by balancing the magnetic field strength against interparticle repulsion[14]. With increasing particle concentration, multiple chains can aggregate into zigzag assemblies, 2D labyrinths, and 3D structures. Here, we found that the magnetic field can be used to drive the interface of a droplet into a highly unstable state by oversaturating the interface with paramagnetic nanoparticle surfactants such that, upon a rapid release of the field, an *explosive* emulsification occurs, where the unstable liquid-liquid interface rapidly generates additional surface area by the formation and ejection of tens of thousands of microdroplets from the interface at high velocities[15-20].

Various mechanisms of spontaneous emulsification have been proposed. Some are mechanical, including interfacial turbulence[21] and interfacial expansion due to a negative interfacial tension[22,23], others are based on phase transformations with the formation of localized areas of supersaturation of surfactants/cosolvents at the interface[24-28]. Still other origins have been proposed, including (a) osmotic pressure gradients[29,30], (b) hydrogen bonding between the surfactants and water increasing the solubility of water in the oil phase[31]; and (c) changes in temperature[18,32], pH[33] or concentration[34]. Here, we report a fundamentally different form of emulsification that is triggered by a high interfacial stress resulting from an oversaturation of the NPSs at the interface by use of an external magnetic field. Due to their strong interfacial activity, if the adsorption of NPSs to a fluid interface reduces the interfacial tension to a low enough value, traditional spontaneous emulsification is observed with the formation of NPS-covered microdroplets that diffuse away from the interface[15]. When the interface is stable, absent an applied field, no instability is observed when a field is applied, even though the interface is oversaturated with NPSs. However, rapidly removing the external field, results in an explosive event in which *ferromagnetic* microdroplets are ballistically ejected from the interface at large ejection velocities. Following this initial explosive emulsification, ferromagnetic microdroplets continue to be released from the parent droplet until the equilibrium NPS surface coverage is restored. Rather than requiring a chemical reaction or meeting specific physiochemical conditions, this unique interfacial response may provide new strategies for the design of micro-ferromagnetic liquid droplet systems[5] and remotely controlled soft microrobots[35,36]. More fundamentally, the use of a magnetic field allows a precise temporal control of the interfacial assembly, enabling a systematic interrogation of interfacial instabilities.

**pH dependence of NPSs assembly**

Carboxylated 30-nm-diameter paramagnetic nanoparticles ($Fe_3O_4$-$CO_2H$) dispersed in an aqueous medium do not assemble at an interface with an apolar solvent, like toluene[37,38]. However, with the addition of the triamine-modified polystyrene (PS-tri$NH_2$), a cationic surfactant, to the oil phase, the protonated PS-tri$NH_3^+$ is able to electrostatically interact with and anchor to the negatively charged $Fe_3O_4$ NPs to form paramagnetic NPSs at the interface, markedly increasing the binding energy of the nanoparticles to the interface[6-8,39-44]. We note that the charge density and binding energy of NPSs are pH-dependent. Under acidic conditions, the amine groups tend to be protonated and the PS-tri$NH_3^+$ exhibits high interfacial activity, which attracts abundant negatively charged NPs to the interface and leads to the formation of NPSs that have a high binding energy and are in a weakly charged state. This could result in a dense packing and even jamming of the NPSs at the interface[6,7]. However, in neutral or alkaline conditions, with fewer amine groups protonated at the interface, less NPSs are formed, and, due to the deprotonation of the carboxyl groups on the NPs' surfaces, there are strong electrostatic repulsions between the NPSs, decreasing the interfacial coverage as evidenced by a higher interfacial tension (Supplementary Fig. 2).

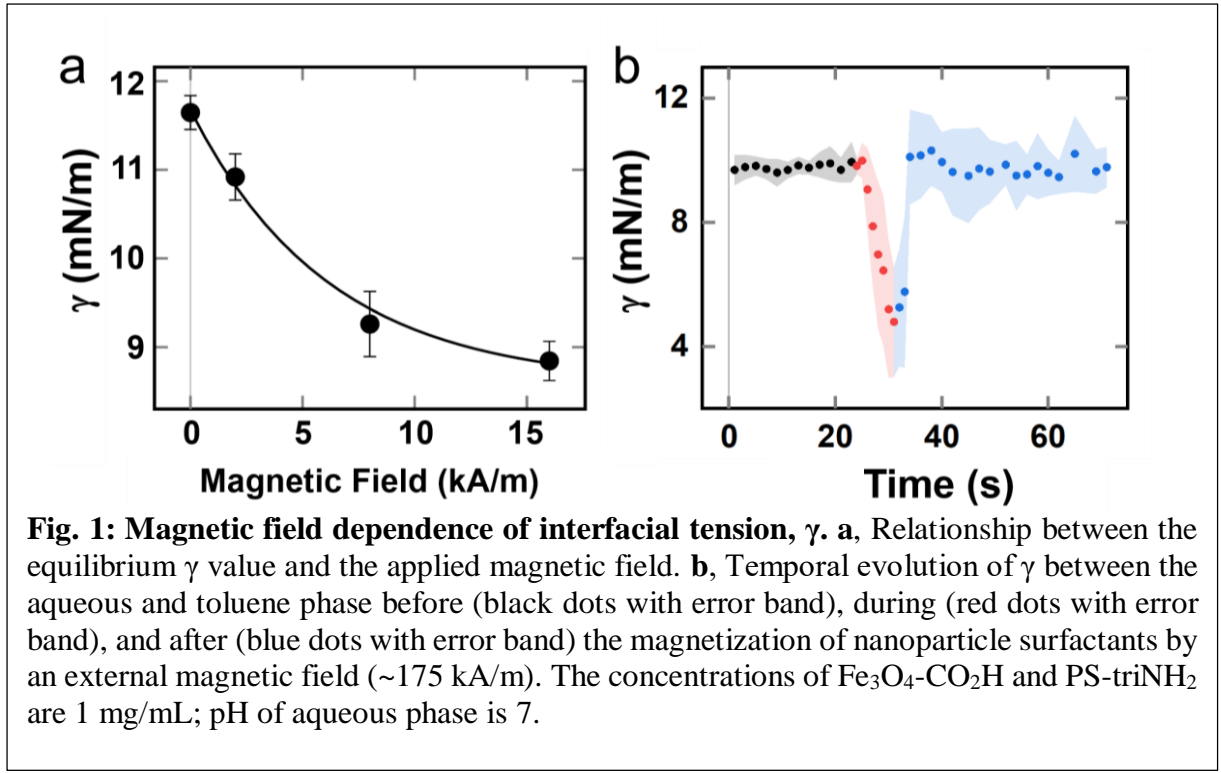

**Fig. 1: Magnetic field dependence of interfacial tension, γ. a**, Relationship between the equilibrium γ value and the applied magnetic field. **b**, Temporal evolution of γ between the aqueous and toluene phase before (black dots with error band), during (red dots with error band), and after (blue dots with error band) the magnetization of nanoparticle surfactants by an external magnetic field (~175 kA/m). The concentrations of $Fe_3O_4$-$CO_2H$ and PS-triNH$_2$ are 1 mg/mL; pH of aqueous phase is 7.

**Magnetic field induced oversaturation**

An external magnetic field induces a magnetic moment in a paramagnetic particle $m = \chi H V$, where $\chi$ is the volume susceptibility of the particle ($Fe_3O_4$), $H$ is the applied magnetic field strength, and $V$ the volume of the particle. When dipolar interactions are strong enough to overcome thermal fluctuations and balance electrostatic repulsions between the particles, the magnetic dipolar force densifies the assembly in a direction along the dipole moment[13,14,45], where the interparticle distance can be tuned with the strength of the magnetic field[14]. A monotonic reduction in the effective interfacial tension is observed with increasing field strength, as shown in Fig. 1a. The effective interfacial tension is the difference between the bare or intrinsic surface tension between the two liquids, $\gamma_{o/w}$, and the in-plane pressure, $P_{2D}$, exerted by the NPSs (i.e., $\gamma = \gamma_{o/w} - P_{2D}$). The significant increase in the in-plane pressure exerted by nanoparticles with the applied field provides compelling indirect evidence for the increased adsorption of NPSs induced by the applied field. The binding energy of the nanoparticles to the interface is governed by the interfacial energy between the liquids and NPSs, material properties that are not altered by the magnetic field.

At neutral/alkaline conditions, where the interfacial packing is dominated by electrostatic repulsions, the application of the external magnetic field can balance the electrostatic repulsion, densify the assembly, and "quench" it into an oversaturated state. In bulk dispersions, this process is usually reversible, and with the field removed, electrostatic repulsion drives the particles away from each other[12,46]. For the interfacial assemblies, though, the binding energy per particle competes with the potential energy after the field is removed. The binding energy of the NPSs, formed by the attachment of multiple hydrophobic ligands to the polar NP, is greatly increased, which inhibits the desorption of the NPs from the interface[6]. As a result, the oversaturated packing of the NPSs is able to raise the in-plane pressure after the field is turned off, triggering the explosive emulsification and rapidly increasing the interfacial tension of the parent drop to its equilibrium value (Fig. 1b). The ejection velocities decrease with increasing pH (Supplementary Fig. 3 and Fig. 2 (filled red circles)).

At low pH and high surfactant concentration, most amine groups are protonated and strongly anchor the surfactants to the $Fe_3O_4$-$CO_2H$. These conditions result in high NPS surface coverage, as indicated by a significantly reduced $\gamma$ (Supplementary Fig. 2) and can ultimately result in spontaneous emulsification where a steady stream of microdroplets flows from the base of the parent droplet (Fig. 2 (filled gold circles) and Supplementary Video 2) [15]. For systems undergoing spontaneous emulsification prior to the application of a field (i.e., low pH and high surfactant concentration), emulsification continues and appears slightly suppressed by the presence of the field (Supplementary Fig. 4). Spontaneous emulsification can also produce a diffuse ring of micro droplets several tens of micrometers away from the parent drop surface (Supplementary Fig. 5) at lower ligand concentrations and pH (Fig. 2 (filled blue

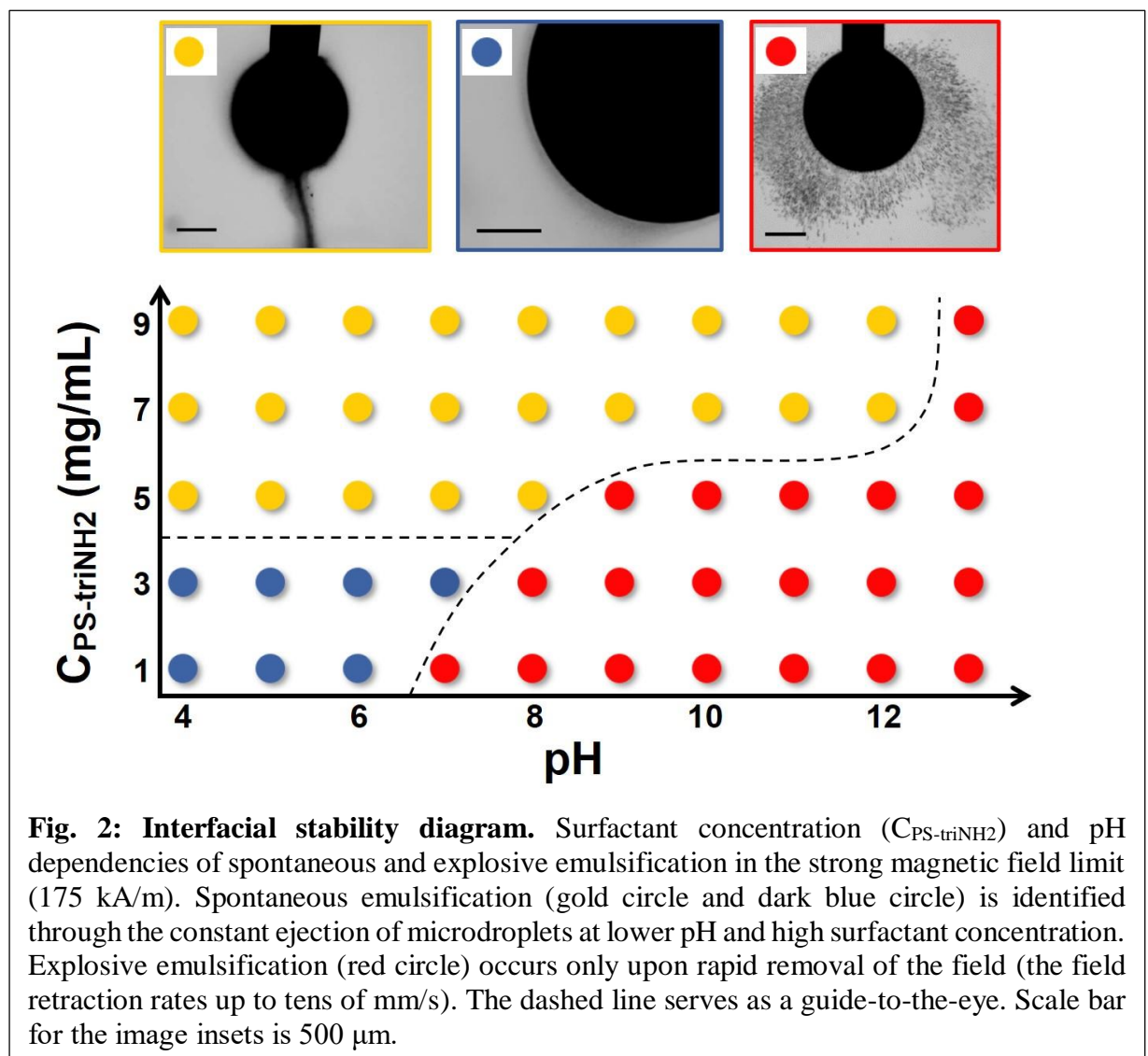

**Fig. 2: Interfacial stability diagram.** Surfactant concentration ($C_{PS-triNH2}$) and pH dependencies of spontaneous and explosive emulsification in the strong magnetic field limit (175 kA/m). Spontaneous emulsification (gold circle and dark blue circle) is identified through the constant ejection of microdroplets at lower pH and high surfactant concentration. Explosive emulsification (red circle) occurs only upon rapid removal of the field (the field retraction rates up to tens of mm/s). The dashed line serves as a guide-to-the-eye. Scale bar for the image insets is 500 μm.

circles)), indicative of an electrostatic repulsion between the parent droplet and the microdroplets. (Supplementary Fig. 3)

**Mechanisms of explosive emulsification**

We leverage computer simulations and continuum theory to reveal the nature of the field-induced driving force for the migration of NPSs to the interface. A representative equilibrium configuration of dipolar magnetic particles confined to the surface of a sphere (representing the parent droplet) in the presence of a uniform magnetic field is obtained from computer simulations (see SI) and depicted in Fig. 3a. The polarization magnitude of nanoparticles is spatially inhomogeneous along the droplet's surface, with a greater degree of polarization along the field direction near the droplet equator. The magnetic field generated by the interfacial dipolar particles is shown in Fig. 3b and reveals significant and nearly uniform magnetic screening of the droplet interior. This screening effect, together with the enhanced polarizability of a particle within the dense shell comprising the interfacial assembly, results in a significant driving force for particle migration to the interface. As shown in Fig. 3c, the chemical potential within the droplet shell (representing the interface) is lower than in the droplet interior. In the framework of linear irreversible thermodynamics, chemical potential gradients drive particle migration[47]. The lower magnetic chemical potential would thus drive particle migration from the droplet interior to the interface until the total chemical potential (which includes contributions from translational entropy and non-magnetic inter-nanoparticle interactions, discussed in our previous work[15]) is spatially uniform. Linear response theory confirms these findings (see SI) and allows us to provide a more direct measure of the driving force for interfacial particle migration through the magnetic chemical potential (Fig. 3c).

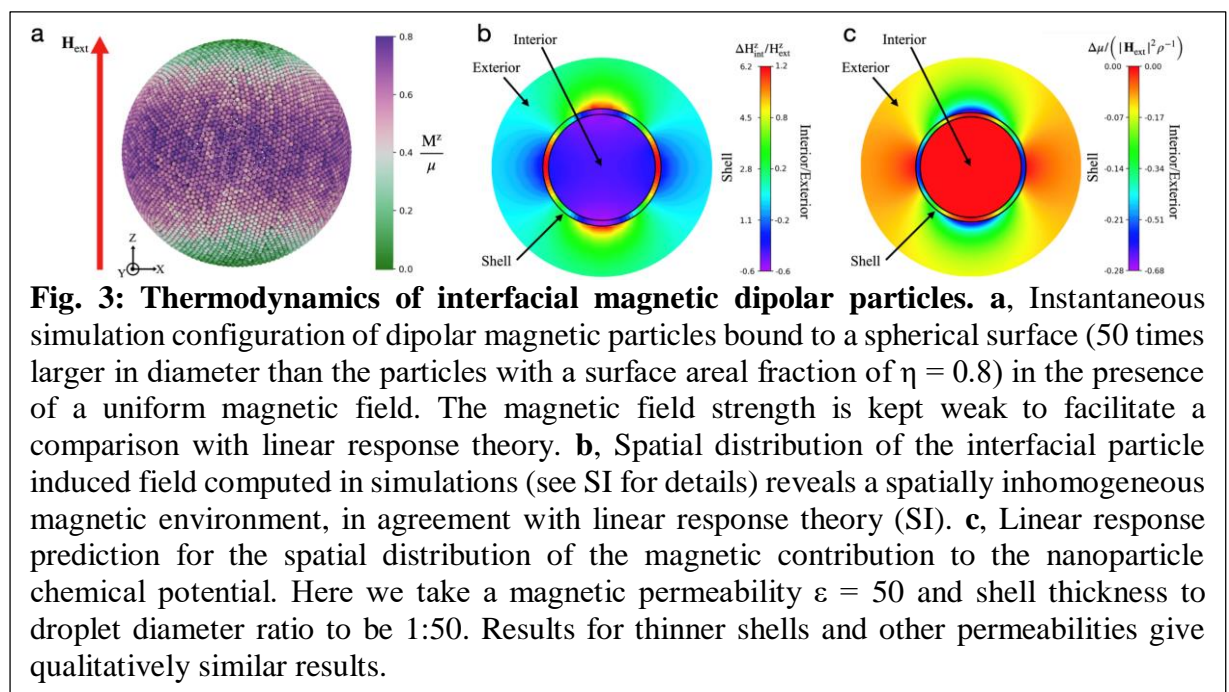

**Fig. 3: Thermodynamics of interfacial magnetic dipolar particles. a**, Instantaneous simulation configuration of dipolar magnetic particles bound to a spherical surface (50 times larger in diameter than the particles with a surface areal fraction of $\eta = 0.8$) in the presence of a uniform magnetic field. The magnetic field strength is kept weak to facilitate a comparison with linear response theory. **b**, Spatial distribution of the interfacial particle induced field computed in simulations (see SI for details) reveals a spatially inhomogeneous magnetic environment, in agreement with linear response theory (SI). **c**, Linear response prediction for the spatial distribution of the magnetic contribution to the nanoparticle chemical potential. Here we take a magnetic permeability $\varepsilon = 50$ and shell thickness to droplet diameter ratio to be 1:50. Results for thinner shells and other permeabilities give qualitatively similar results.

Despite the enhanced NPS surface coverage resulting in what can be up to an order of magnitude decrease in the $\gamma$ from the pre-field value, the aqueous droplets remain stable while the field is present. For systems that were stable but near the emulsification boundary in the absence of the field, one might expect that the field-induced increase in interfacial NPS coverage would provide the requisite driving force for emulsification. Nevertheless, emulsification was not observed despite the observed reduction of $\gamma$ and the proximity to the emulsification boundary.

The apparent stability of these interfaces has several possible origins. One is that the interface, despite its high NPS coverage, is fully equilibrated and thus truly stable. The stability criteria for an interface are sensitive to the in-plane interactions between interfacial NPSs (or more directly the in-plane pressure, $P_{2D}$, exerted by the NPSs)[15]. The magnetic field acts to polarize the magnetic dipoles in the field-direction, which in turn alters the nanoparticle microstructure. While this undoubtedly affects the in-plane pressure tensor that is decreased under the magnetic field owing to the existence of dipolar-dipolar interaction based upon the Frumkin equation[48,49], the degree to which this occurs is presently unclear. An alternative and intriguing possibility is that the droplet is metastable due to the NPS microstructure introducing a barrier to emulsification, the globally stable system configuration. As a well-established example of such metastability, liquid-liquid interfaces that are maximally packed or "jammed" with NPSs can persist indefinitely and exhibit solid-like properties[5-7,40]. A hallmark of these structured liquids is the development of an elastic response to interfacial deformation. In the present case of magnetic NPSs, however, the surface coverage is below the threshold for jamming as determined from wrinkling experiments (Supplementary Fig. 6), where a wrinkle pattern is not observed on the pendant droplet surface during a reduction of the droplet volume[7].

**Magnet retraction rate dependence**

Under conditions where the interface is stable in the absence of a field, application and removal of the magnetic field can lead to different scenarios depending on the field retraction rate. Removing the field introduces a thermodynamic driving force for the desorption of the excess NPSs back into the aqueous parent droplet until the equilibrium coverage is restored. As a result, removing the field quasi-statically (i.e., reversibly removing the external field) should result in the steady desorption of NPS into the aqueous phase and a return to stable interface.

If the field is instead removed at a finite rate, nonequilibrium dynamical effects may dominate the response. Since the degree of paramagnetic NPS adsorption increases with the magnetic field strength, we focus on strong magnetic field strengths (~175 kA/m) with a magnetization time of ~10 s before retracting the field (Supplementary Fig. 3e) (see Methods). Rapidly removing the external magnetic field allows for probing the irreversible limit and observing a forceful ejection of microdroplets after a short delay, i.e., within two seconds after removal of the field (Supplementary Fig. 3 and Supplementary Fig. 7).

At the fastest removal rate, tens of thousands of microdroplets (Fig. 4a), ~4 μm in diameter, are ejected at a velocity of ~3.7 mm/s (Fig. 4b) and continue their persistent motion over 1-2 mm away from the parent droplet surface. This explosive event stabilizes the interface, reflected in the restoration of $\gamma$ to its pre-field values 3 seconds after removing the field (Fig. 1b). Decreasing the field removal rate monotonically reduces the number of ejected microdroplets and their ejection velocities, and, also, increases the delay time between field removal and the explosive event (Fig. 4). The energy released through explosive emulsification is decreased at slower removal rate, where the remainder of the excess energy is dissipated by particle rearrangement and desorption from the interface. The diminishment of this explosive event with decreasing field removal rate is consistent with our expectation of no explosive behavior in the quasi-static limit.

While the ejection velocities of these microdroplets can be extraordinarily high, the colloidal scale of these droplets and the viscosity of the oil would suggest that low Reynolds number physics is at play and inertial effects are negligible. The persistent motion for these large length scales is therefore likely the result of a persistent force acting on the microdroplets. Magnetic dipolar forces between the parent and microdroplets could provide an initial ejective force, but the ejected microdroplets would rapidly reorient to alleviate such a repulsion. The long-range nature of the force apparently propelling the microdroplets is highly suggestive of an electrostatic origin, which may be the result of overcharging the interface upon field removal. The magnetic field brings excess negatively charged $Fe_3O_4$ NP to the interface, which, we hypothesize, is similar to overcharging the interface with negative charges. When the field is on, the dipolar-dipolar interaction between the particles can stabilize the droplet in the overcharged state, which can return to equilibrium state by propelling the excess nanoparticles from the interface by explosive emulsification. Imaging experiments using upconverting nanoparticles,[50,51] whose emission is sensitive to sub-degree temperature changes,[52] did not show measurable temperature changes during the explosive emulsification, arguing against a latent heat origin of the observed behavior.

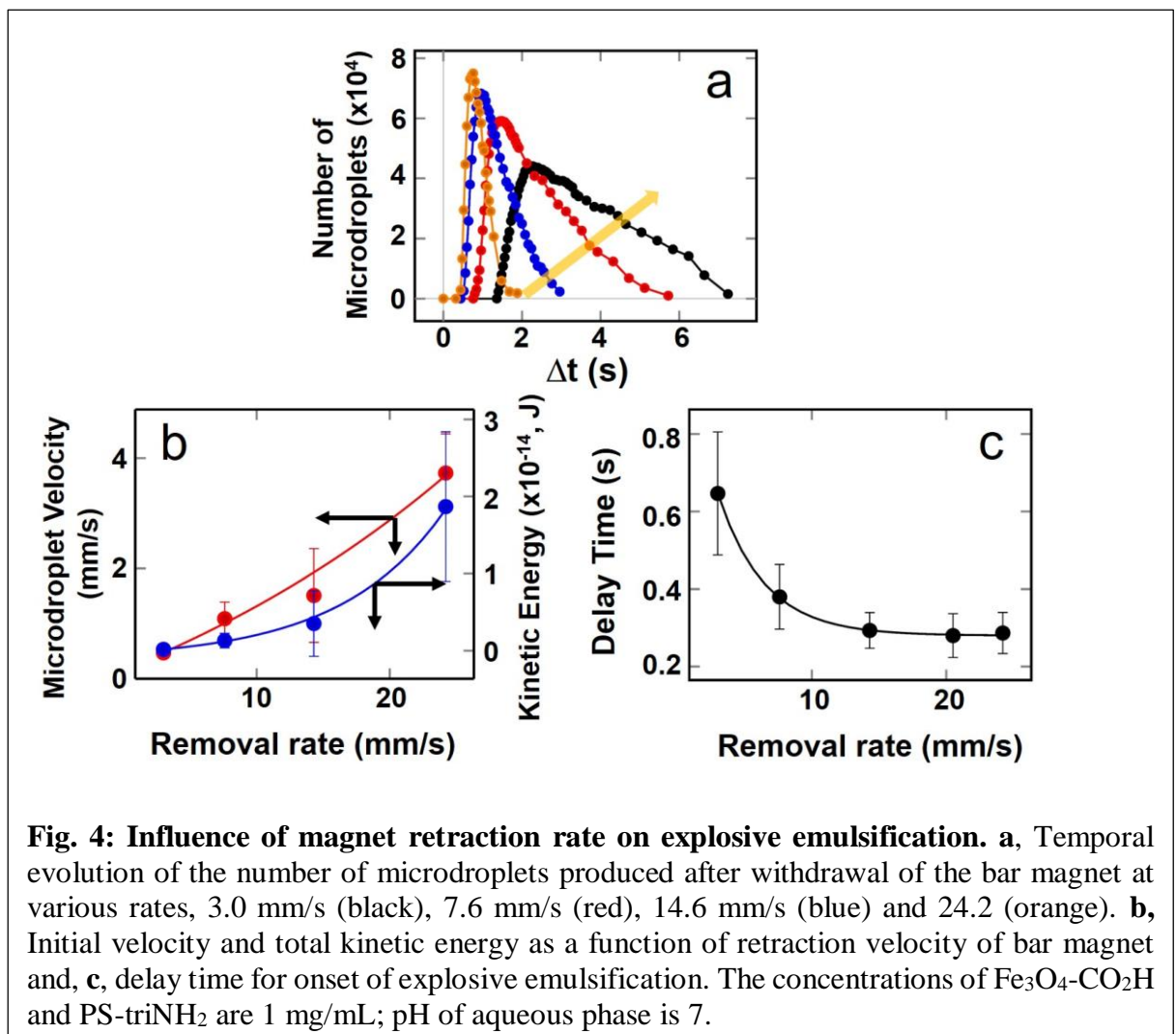

**Fig. 4: Influence of magnet retraction rate on explosive emulsification. a,** Temporal evolution of the number of microdroplets produced after withdrawal of the bar magnet at various rates, 3.0 mm/s (black), 7.6 mm/s (red), 14.6 mm/s (blue) and 24.2 (orange). **b,** Initial velocity and total kinetic energy as a function of retraction velocity of bar magnet and, **c,** delay time for onset of explosive emulsification. The concentrations of $Fe_3O_4$-$CO_2H$ and PS-tri$NH_2$ are 1 mg/mL; pH of aqueous phase is 7.

**Characterization of the microdroplets**

The microdroplets produced during explosive emulsification and also spontaneous emulsification have a remnant magnetization. The remnant magnetization of microdroplets with an average diameter of ~4 μm is found to be ~$1.7 \times 10^{-16}$ A.m$^2$, calculated using the methods mentioned in ref. 41, and a sizable coercive field (ferromagnetism)[41,53] as verified by their rotation in the presence of an external rotating magnetic field (Supplementary Video 3 and Fig. 5b). Previously, the development of ferromagnetism was associated with the jamming of paramagnetic NPSs at the interface.[5,41] This raises the intriguing possibility that despite the parent droplets being decidedly below the jamming transition, the microdroplets generated during explosive emulsification may have a NPS surface coverage that exceeds that of the parent droplet. In the long-time limit and in the absence of a field, these droplets should eventually merge with the parent droplet. The stability as evidenced by the long lifetime of the microdroplets, along with their ferromagnetism, supports the conclusion that they are in fact jammed (Fig. 5b, c).

As noted above, high surfactant concentrations significantly reduce γ and promote, in the absence of a magnetic field, spontaneous emulsification. This is evidenced by the microdroplets streaming off the base of the drop under the influence of gravity. The shape of the stream is highly dependent on pH (Fig. 5a). At low pH, the stream bows upward as the microdroplets leave the drop, and, falls straight downwards at high pH. This difference arises from the difference in the size of the microdroplets (as determined by light scattering) (Fig. 5c and Supplementary Fig. 8). Although the density of microdroplets is close to the density of water (1 g/cm$^3$) and, therefore, greater than that of the surrounding medium (0.867 g/cm$^3$), Brownian forces and convection (driven by small temperature gradients) become comparable to gravitational force, and influence the shape of the stream.

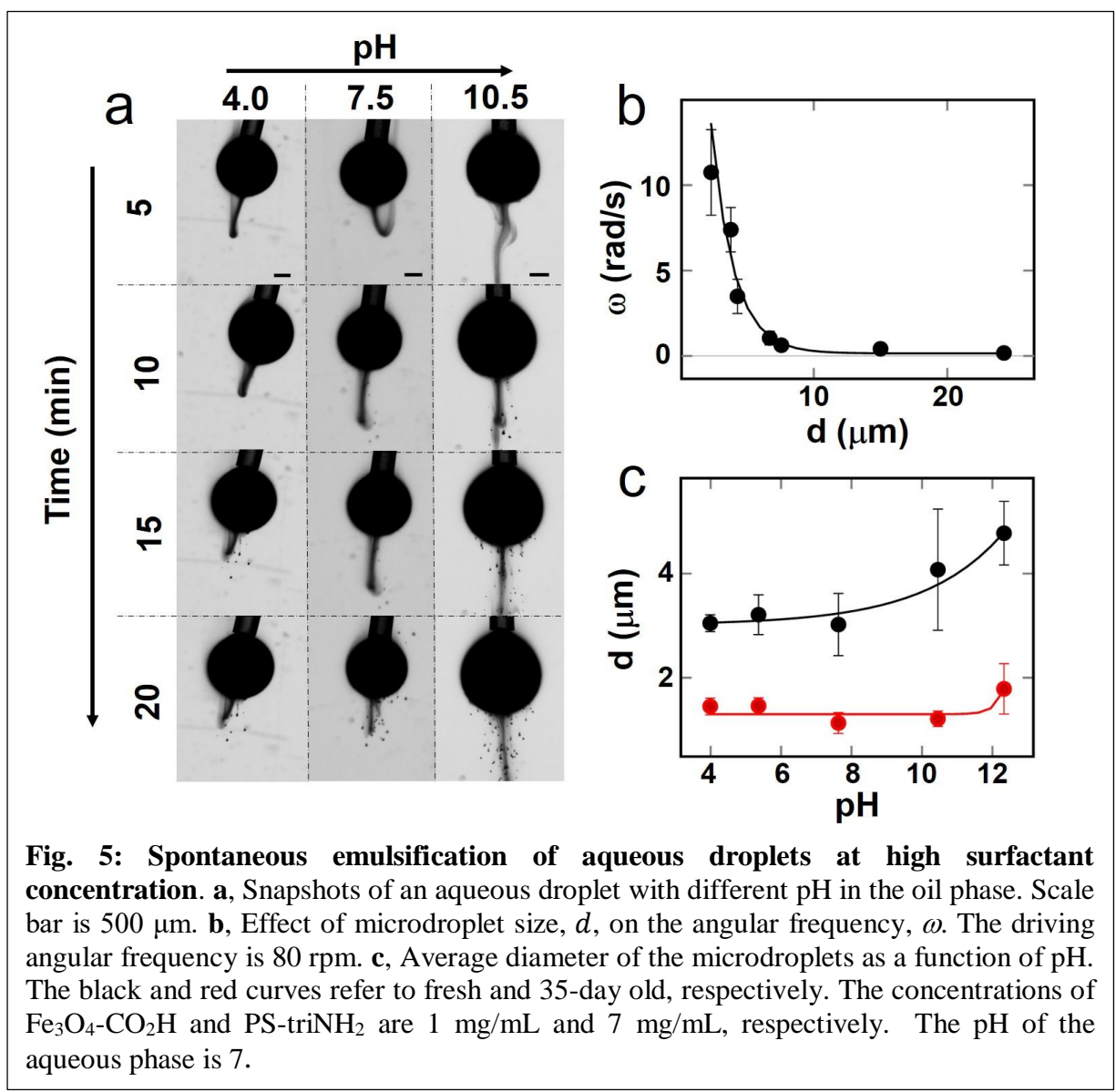

**Fig. 5: Spontaneous emulsification of aqueous droplets at high surfactant concentration.** **a**, Snapshots of an aqueous droplet with different pH in the oil phase. Scale bar is 500 μm. **b**, Effect of microdroplet size, $d$, on the angular frequency, $\omega$. The driving angular frequency is 80 rpm. **c**, Average diameter of the microdroplets as a function of pH. The black and red curves refer to fresh and 35-day old, respectively. The concentrations of $Fe_3O_4$-$CO_2H$ and PS-triNH$_2$ are 1 mg/mL and 7 mg/mL, respectively. The pH of the aqueous phase is 7.

An optical image of the microdroplets suspended in the oil phase is shown in Supplementary Fig. 9. Except for large droplets that sediment due to gravity (Supplementary Fig. 10), the microdroplets are stable for at least one month without coalescence or loss of their magnetization. After this time, the microdroplets shrink to half their original size (Fig. 5c).

**Conclusions**

In conclusion, assemblies of paramagnetic NPS at the oil/water interface were brought into an oversaturated using an external dc magnetic field. Upon rapid removal of the field, the stored free energy is released through an *explosive* emulsification, a dynamical interfacial instability where an explosive ejection of a plume of ferromagnetic microdroplets from the interface occurs. The potential energy of the parent drop and ejection velocities of the microdroplets were tunable by changing the pH, surfactant concentration, and rate of removal of the permanent magnet from the drop. These findings uncover an efficient energy storage and release process, that has potential applications for remotely controlled soft microrobots, taking advantage of the ferromagnetic nature of the microdroplets. The controllable mass transfer during the explosive emulsification also endows more possibilities to realize controllable and intermittent cargo delivery, chemical shuttling in compartmentalized micro-reaction systems and liquid external field sensors.

**Methods**

**Preparation of paramagnetic nanoparticle surfactants at the interface.** The negatively charged carboxylic acid-functionalized iron oxide nanoparticles ($Fe_3O_4$-$CO_2H$) (Ocean NanoTech) with the diameter of ~30 nm are superparamagnetic nanoparticles with excellent colloidal stability. One monolayer of oleic acid and one monolayer of amphiphilic polymer, totaling ~4 nm, surrounded the ~22 nm-sized magnetic core. They are prepared by thermos-decomposition method and each nanoparticle is a single crystal with a structure of magnetite. The $Fe_3O_4$-$CO_2H$ nanoparticles were dispersed in deionized water to form the aqueous dispersion that was injected into the oil phase by the syringe to form a pendant liquid droplet.

The ω-(diethylene triamine)-terminated polystyrene (PS-triNH$_2$, Mw = 1200 g/mol) (Polymer Source) was dissolved in toluene to interact with $Fe_3O_4$-$CO_2H$ nanoparticles, forming the magnetic NPSs at the interface. The polydispersity index (PDI) for the PS-triNH$_2$ is 1.25, and the functionality is 98%. The $Fe_3O_4$-$CO_2H$ concentration remained 1 mg/mL; the concentration of PS-triNH$_2$ ranged from 1 to 9 mg/mL. All nanoparticle dispersions were used without further purification and diluted to the required concentration using deionized water. The pH of the aqueous dispersions was adjusted using 1.0 M NaOH or HCl.

**Interfacial tension measurement.** A tensiometer (Krüss DSA30) was used to measure γ between water and toluene via the pendant drop method. The time dependence of γ was recorded after the aqueous droplet had been injected into the oil phase. The volume of the aqueous droplet was ~5 μL, and the measurement time was up to 3000 s. A homemade Helmholtz coil was used to magnetize the magnetic NPSs in a uniform magnetic field up to 16 kA/m. The droplet volume remained constant in the presence of the magnetic field, indicating that the heat generated by the coil and its effect on the assembly were negligible. A NdFeB magnet was also used to magnetize the magnetic NPSs (~175 kA/m), and 2D simulation of the magnetic field surrounding the magnet is shown in Supplementary Fig. 11 (COMSOL Multiphysics).

**Optical observation of the interfacial behavior.** A digital camera on a tensiometer was used to visualize the macroscopic behavior of the magnetic NPSs at and near the interface, with the pH of aqueous phase and surfactant concentration varied. A strong bar magnet (NdFeB) was used to magnetize the droplet; the magnetic field strength and direction were the same for all measurements (Supplementary Fig. 11). The droplets were exposed for 10 s to a magnetic field of ~175 kA/m.

**Microscopic characterization of the microdroplets.** The micromorphology of the microdroplets was characterized by polarized optical microscopy (ZEISS Imager.A2). The diameter of the microdroplet suspended in toluene was detected by means of dynamic light scattering (Malvern Zetasizer Nano Series ZS90), and the microdroplet dispersions were prepared by spontaneously forming microdroplets at the interface of the hanging aqueous droplet in the surrounding oil phase for 3 h. The upper layer of the microdroplet dispersion was used to measure the microdroplet size without additional procedures.

**Controllable retraction rate of the magnet.** A homebuilt stepping motor was used to control the speed of retraction of a NdFeB magnet that was attached to a translation stage. The motor speed was varied up to 1000 rpm equivalent to a velocity of 3 to 24.2 mm/s. The magnet was mechanically fixed to the stage to ensure the same magnetic field strength and direction for all measurements (Supplementary Fig. 11).

**Statistical Analysis** All reported experiments in this study were reproduced in triplicate to confirm data consistenc. The results were expressed as mean ± standard deviation (SD). Python, ImageJ and OriginPro software were used to analyze, visualize, and plot all data.

**Data Availability**

All data supporting the findings of this study are available within the Article and the Supplementary Information, and from the corresponding author on reasonable request.

**Acknowledgements**

The experiments were supported by the U.S. Department of Energy, Office of Science, Office of Basic Energy Sciences, Materials Sciences and Engineering Division under Contract No. DE-AC02-05-CH11231 within the Adaptive Interfacial Assemblies Towards Structuring Liquids program (KCTR16). The authors also acknowledge the support of the Beijing Advanced Innovation Center for Soft Matter Science and Engineering at the Beijing University of Chemical Technology. R.S. acknowledges support from the National Science Foundation (NSF), Division of Materials Research (DMR) under Grant No. 2203933. Work at the Molecular Foundry was supported by the Director, Office of Science, Office of Basic Energy Sciences, Division of Materials Sciences and Engineering, of the U.S. Department of Energy under Contract No. DE-AC02-05CH11231.

**Author contribution**

X.W., D.W. and T.P.R. designed the experiments. X.W., H.X., C.C.S.P., and B.R. performed experiments. G.B., R.S., J.H., A.K.O., P.L.G., B.E.C. and P.A. developed the theoretical background and performed the simulations. X.W., G.B., R.S., J.H., A.K.O., P.L.G., D.W., H.X., J.W., C.C.S.P., B.E.C. and P.A. and T.P.R. had discussions on the results and analysis. X.W., A.K.O, P.L.G. and T. P. R. drafted the manuscript, and all the authors were involved in the discussion of results and the final manuscript editing.

**Corresponding authors**

Correspondence to Ahmad K. Omar, Dong Wang or Thomas P. Russell

**Competing interests**

The authors declare no competing interests.

# Supplementary Materials for

**Ballistic Ejection of Microdroplets from Overpacked Interfacial Assemblies**

Xuefei Wu, Gautam Bordia, Robert Streubel, Jaffar Hasnain, Cássio C.S. Pedroso, Bruce E. Cohen, Behzad Rad, Paul Ashby, Ahmad K. Omar*, Phillip L. Geissler, Dong Wang*, Han Xue, Jianjun Wang, Thomas P. Russell*

Correspondence to: aomar@berkeley.edu; dwang@mail.buct.edu.cn; russell@mail.pse.umass.edu

**This PDF file includes:**

Supplementary Text
Supplementary Figs. 1 to 11
Captions for Supplementary Videos 1 to 3

**Other Supplementary Materials for this manuscript include the following:**

Supplementary Videos 1 to 3

**Supplementary Text**

To estimate the response of interfacial nanoparticles (NPs) to an external magnetic field, we have: (i) solved magnetostatic continuum equations within a linear response approximation, and; (ii) performed computer simulations of interacting magnetic particles on the surface of a sphere. Results of these two calculations, which agree qualitatively, are shown in Fig. 3 of the main text. Details of these calculations are presented here in addition to further analysis.

<u>Linear Response Theory</u>

An isolated paramagnetic NP (or a freely rotating superparamagnetic NP) will develop a nonzero average magnetic dipole $\langle \mathbf{M} \rangle_{\mathbf{H}_{ext}}$ in the presence of an external magnetic field $\mathbf{H}_{ext}$. The expected linear growth of $|\langle \mathbf{M} \rangle_{\mathbf{H}_{ext}}|$ with $|\mathbf{H}_{ext}|$ can be described by an effective free energy

$$F(\mathbf{M}) = \frac{1}{2\alpha} |\mathbf{M}|^2 - \mathbf{H}_{ext} \cdot \mathbf{M}$$

with polarizability $\alpha = \beta \langle |\mathbf{M}|^2 \rangle_0$, where $\beta = (k_B T)^{-1}$ and $T$ is temperature. The chemical potential associated with this particle's dipole fluctuations is

$$\mu = k_B T \ln \int d\mathbf{M} e^{-\beta F(\mathbf{M})} = \text{const} - \frac{\alpha}{2} |\mathbf{H}_{ext}|^2$$

where the constant is independent of $\mathbf{H}_{ext}$. We will focus exclusively on changes in $\mu$ due to the external field, $\Delta\mu = \mu(\mathbf{H}_{ext}) - \mu(0)$. For the isolated NP, we then have

$$\Delta\mu = -\frac{\alpha}{2} |\mathbf{H}_{ext}|^2 \tag{S1}$$

A dense collection of such particles, within a volume $V$, can be viewed on sufficiently large length scales as a polarizable continuum with free energy

$$F[\mathbf{m}(\mathbf{r})] = \frac{1}{2} \int d\mathbf{r} \int d\mathbf{r}' \, \mathbf{m}(\mathbf{r}) \cdot \left[ \left((\rho\alpha)^{-1} - \frac{4\pi}{3}\right) \mathbf{I} \delta(\mathbf{r} - \mathbf{r}') + \nabla\nabla' \frac{1}{|\mathbf{r}-\mathbf{r}'|} \right] \cdot \mathbf{m}(\mathbf{r}') - \int d\mathbf{r} \mathbf{H}_{ext} \cdot \mathbf{m}(\mathbf{r}) \tag{S2}$$

where $\rho$ is the number density of NPs and $\mathbf{m}(\mathbf{r})$ is the net dipole per unit volume in a small region at position $\mathbf{r}$. The dipole-dipole interaction tensor $\nabla\nabla' |\mathbf{r}-\mathbf{r}'|^{-1}$ gives a singular but integrable self-interaction at $\mathbf{r} = \mathbf{r}'$, whose contribution is subtracted in the first term of Eq. S2. The average behavior of this linear response model is equivalent [1-3] to solutions of

$$\nabla \cdot \mathbf{H} = 0 \tag{S3}$$

where

$$\mathbf{H} = \mathbf{H}_{ext} - \int d\mathbf{r}' \nabla\nabla' \frac{1}{|\mathbf{r}-\mathbf{r}'|} \cdot \mathbf{m}(\mathbf{r}') \tag{S4}$$

is the total magnetic field at $\mathbf{r}$, accompanied by the linear response relation

$$\mathbf{m}(\mathbf{r}) = \frac{\varepsilon - 1}{4\pi} \mathbf{H}(\mathbf{r})$$

The relative magnetic permeability $\varepsilon$ is determined by microscopic parameters through the magnetic analog of the Clausius-Mossotti equation, $4\pi\rho\alpha/3 = (\varepsilon-1)/(\varepsilon+2)$. (This permeability is conventionally denoted $\mu$, a symbol we reserve for chemical potential. The $\varepsilon$-notation is a reminder that the entire linear response calculation would proceed identically for a collection of electric dipoles, for which $\varepsilon$ represents the dielectric constant.)

Boundary conditions must be satisfied at interfaces between $V$ and regions $\tilde{V}$ lacking NPs

$$\varepsilon \mathbf{H}_V \cdot \hat{\mathbf{n}} = \mathbf{H}_{\tilde{V}} \cdot \hat{\mathbf{n}} \tag{S5}$$

$$\mathbf{H}_V \cdot \hat{\mathbf{t}} = \mathbf{H}_{\tilde{V}} \cdot \hat{\mathbf{t}} \tag{S6}$$

$\hat{\mathbf{n}}$ and $\hat{\mathbf{t}}$ are unit vectors pointing normal and tangential, respectively, to the local surface.

As an idealization of paramagnetic NPs adsorbed at a droplet interface, we consider a polarizable spherical shell, with inner radius $a$ and outer radius $b$, centered at the origin (see

Supplementary Fig. 1). We take the external magnetic field, $\mathbf{H}_{ext} = H_{ext}\hat{\mathbf{z}} = $ const, to be uniform and oriented along the polar direction $\hat{\mathbf{z}}$.

For a time-independent external field, $\nabla \times \mathbf{H} = 0$, so that $\mathbf{H}$ can be written as the gradient of a scalar potential,
$$\mathbf{H}(\mathbf{r}) = -\nabla \phi(\mathbf{r})$$
We must therefore solve a Laplace equation,
$$\nabla^2 \phi(\mathbf{r}) = 0$$
within each of three regions, namely "in" ($r < a$), "shell" ($a \leq r \leq b$), and "out" ($r > b$). The boundary conditions (S6) are supplemented by the requirement that $\mathbf{H}(\mathbf{r})$ asymptotically approaches $\mathbf{H}_{ext}$ at large $r$.

The axial symmetry of this problem permits a general solution to the Laplace equation in terms of $r$ and the polar angle $\theta = \cos^{-1}(\hat{\mathbf{z}} \cdot \hat{\mathbf{r}})$,
$$\phi(r, \cos\theta) = \sum_{l=0}^{\infty}[B_l r^l + C_l r^{-(l+1)}]P_l(\cos\theta)$$
where $P_l(x)$ are Legendre polynomials. $B_l$ and $C_l$ are constant within each of the three regions but may change discontinuously at the boundaries between them. For $l = 0$ and $l > 1$, boundary conditions are satisfied with $B_l = C_l = 0$ and so we will henceforth omit the subscript $l$ with $l = 1$ implied. The solution to this boundary value problem is then determined by the six remaining coefficients, i.e., values of $B$ and $C$ in the exterior, shell, and interior regions:

$$B^{shell} = -3|\mathbf{H}_{ext}|\left[\varepsilon + 2 - 2\left(\frac{a}{b}\right)^3 \frac{(\varepsilon-1)^2}{2\varepsilon+1}\right]^{-1} \quad (S7)$$

$$C^{shell} = a^3 \frac{\varepsilon-1}{2\varepsilon+1} B^{shell} \quad (S8)$$

$$B^{in} = B^{shell} + C^{shell} a^{-3} = B^{shell}\left(1 + \frac{\varepsilon-1}{2\varepsilon+1}\right) \quad (S9)$$

$$C^{in} = 0 \quad (S10)$$

$$C^{out} = b^3[\varepsilon_0 + B^{shell} + C^{shell} b^{-3}] \quad (S11)$$

$$B^{out} = -H_{ext} \quad (S12)$$

The resulting total magnetic field follows by differentiation:
$$\mathbf{H} = -\nabla[Br + Cr^{-2}\cos\theta]$$
Giving

$$\mathbf{H}^{in} = -B^{in}\hat{\mathbf{z}} \quad (r < a) \quad (S13)$$

$$\mathbf{H}^{shell} = B^{shell}\hat{\mathbf{z}} - \frac{C^{shell}}{r^3}\left(\hat{\mathbf{z}} - 3\frac{z}{r}\hat{\mathbf{r}}\right) \quad (a < r < b) \quad (S14)$$

$$\mathbf{H}^{out} = H_{ext}\hat{\mathbf{z}} - \frac{C^{out}}{r^3}\left(\hat{\mathbf{z}} - 3\frac{z}{r}\hat{\mathbf{r}}\right) \quad (r > b) \quad (S15)$$

In the droplet's interior ($r < a$), the external field is uniformly screened by the polarized shell, resulting in a uniform total field (Eq. S13). The field-induced chemical potential shift is therefore also spatially uniform. Adapting Eq. S1, we have
$$\Delta\mu^{in} = -\frac{\alpha}{2}|\mathbf{H}^{in}|^2 = -\frac{1}{2}\left(\frac{\varepsilon-1}{4\pi\rho}\right)\left(\frac{3}{\varepsilon+2}\right)|\mathbf{H}^{in}|^2$$

From similar considerations, the dilute region outside the shell has a spatially varying field (Eq. S14) and therefore a spatially varying chemical potential
$$\Delta\mu^{out}(\mathbf{r}) = -\frac{1}{2}\left(\frac{\varepsilon-1}{4\pi\rho}\right)\left(\frac{3}{\varepsilon+2}\right)|\mathbf{H}^{out}(\mathbf{r})|^2$$

To compute $\Delta\mu$ for $a < r < b$, we must account for the self-interaction implicit in the continuum expression (Eq. S4). Identifying a sub-volume of size $\rho^{-1}$ as the continuum equivalent of a single particle, we use Eq. S1 with re-normalized $\alpha$ and $\mathbf{H}$ such that $\bar{\alpha} = (\varepsilon - 1)(2\varepsilon + 1)/(12\pi\rho\varepsilon)$ and $\bar{\mathbf{H}}(\mathbf{r}) = 3\varepsilon/(2\varepsilon + 1)\mathbf{H}(\mathbf{r})$ [4]

$$\Delta\mu^{shell}(\mathbf{r}) = -\frac{\bar{\alpha}}{2}|\bar{\mathbf{H}}^{shell}|^2 = -\frac{1}{2}\frac{(\varepsilon-1)(2\varepsilon+1)}{(12\pi\rho\varepsilon)}\left(\frac{3\varepsilon}{2\varepsilon+1}|\mathbf{H}^{shell}(\mathbf{r})|\right)^2 \tag{S16}$$

Results of this analysis, shown in Fig. 3c, suggest that paramagnetic NPs experience a chemical potential gradient under an applied magnetic field, driving material away from the screened droplet interior and from the poles of the shell, towards the shell's equator, and towards the polar region of the droplet's exterior. Experimentally, the nanoparticle shell thickness is vanishingly small in comparison to the droplet radius. Upon taking the thin shell limit ($a \to b$) we find the external field is no longer screened by the shell except with-in the shell:

$$\mathbf{H}^{in} = \mathbf{H}^{out} = \mathbf{H}_{ext} \tag{S17}$$

$$\mathbf{H}^{shell} = \frac{|\mathbf{H}_{ext}|}{3\varepsilon}\left[(2\varepsilon+1)\hat{\mathbf{z}} + (\varepsilon-1)\frac{a^3}{r^3}\left(\hat{\mathbf{z}} - 3\frac{z}{r}\hat{\mathbf{r}}\right)\right] \tag{S18}$$

$$\hat{\mathbf{z}} \cdot \mathbf{H}^{shell} = |\mathbf{H}_{ext}|\left[1 - \frac{(\varepsilon-1)}{\varepsilon}\cos^2\theta\right] \tag{S19}$$

In this limit the chemical potential is spatially uniform except within the shell, where it varies greatly from pole to equator. At the equator $\mu$ is much lower than in solution:

$$\Delta\mu^{shell}_{equator} = \Delta\mu^{in}\,\varepsilon\left(\frac{\varepsilon+2}{2\varepsilon+1}\right) < \Delta\mu^{in} \tag{S20}$$

By contrast, at the pole $\mu$ is higher than in solution:

$$\Delta\mu^{shell}_{pole} = \Delta\mu^{in}\,\frac{1}{\varepsilon}\left(\frac{\varepsilon+2}{2\varepsilon+1}\right) > \Delta\mu^{in} \tag{S21}$$

Among the approximations we have made, treating a monolayer of NPs as a continuous field is likely the most aggressive. It is also straightforward to scrutinize with computer simulations of systems comprising discrete NPs, as described in the next section.

Simulations

To verify the predictions from the continuum theory analysis we conduct molecular dynamics simulations of dipolar magnetic nanoparticles constrained to the surface of a sphere. We simulate $N$ dipolar particles in a cubic box with volume $V$. Dipolar particles interact with both an effective hard-sphere-like potential and a long-ranged dipolar potential and are also subject to an external magnetic field $\mathbf{H}_{ext}$.

A liquid droplet is represented as a sphere of radius $R$, centered within the simulation volume, to which particles are bound. More precisely, particles are subject to a potential that is harmonic in radial displacements away from the interface. The potential is sufficiently deep such that all particles irreversibly adsorb onto the interface. The total potential energy of the system $U$ is given by the sum of dipolar $U_{dp}$, volume exclusion $U_{ve}$, external $U_{ext}$, and interfacial binding $U_{int}$ potentials with these interactions taking the following functional forms:

$$U = U_{dp} + U_{ve} + U_{ext} + U_{int} \tag{S22}$$

$$U_{dp} = \frac{1}{2}\sum_{i,j\neq i}\frac{1}{r^3}(\mathbf{m}_i \cdot \mathbf{m}_j) - \frac{3}{r^5}(\mathbf{m}_i \cdot \mathbf{r})(\mathbf{m}_j \cdot \mathbf{r}) \tag{S23}$$

$$U_{ve} = \frac{1}{2}\sum_{i,j\neq i}4\varepsilon\left[\left(\frac{\sigma}{r}\right)^{12} - \left(\frac{\sigma}{r}\right)^{6} - \left(\frac{\sigma}{r_c}\right)^{12} + \left(\frac{\sigma}{r_c}\right)^{6}\right] \tag{S24}$$

$$U_{ext} = -\sum_i \mathbf{m}_i \cdot \mathbf{H}_{ext} \tag{S25}$$

$$U_{int} = \sum_i \kappa(r_i - R_{int})^2 \tag{S26}$$

where $\mathbf{r} = \mathbf{r_i} - \mathbf{r_j}$ is the displacement vector (with magnitude $r$) between particles $i$ and $j$ and $\mathbf{m_i}$ is the dipole vector of particle $i$ with constant magnitude $m$ such that $\mathbf{m_i} = m\mathbf{q_i}$ where $\mathbf{q_i}$ is a unit vector. Volume exclusion is enforced by a hard-sphere-like interaction, implemented using a stiff Weeks-Chandler-Anderson (WCA) potential [5], where $\varepsilon$ is the Lennard Jones (LJ) interaction energy, $\sigma$ is the LJ diameter (all references made to the particle diameter refer to the LJ diameter) and $r_c = 2^{1/6}\sigma$ is the cutoff at which the potential is truncated. By setting $\varepsilon = 100 k_B T$, we ensure nearly complete volume exclusion with an effective hard-sphere diameter of approximately $r_c$. Finally, particles are constrained to the interface by choosing a sufficiently large binding energy ($\kappa = 1000 k_B T/\sigma^2$) such that the magnitude of the particle's position vector $r_i$ (relative to the origin of the droplet) is equal to the droplet's radius $R_{\text{int}}$. This choice results in all particles in the simulation cell adsorbing to the droplet surface.

Brownian dynamics simulations were used to evolve our system forward in time towards the stationary state. The corresponding translational and rotational equations-of-motion for a particle $i$ are given by

$$m\ddot{\mathbf{r}}_i = \boldsymbol{\xi}_i^T - \frac{\partial U}{\partial \mathbf{r}_i} - \gamma^T \dot{\mathbf{r}}_i \tag{S27}$$

$$I\dot{\boldsymbol{\omega}}_i = \boldsymbol{\xi}_i^R - \frac{\partial U}{\partial \mathbf{q}_i} - \gamma^R \boldsymbol{\omega}_i \tag{S28}$$

where $\gamma^T$ and $\gamma^R$ are the translational and rotational drag coefficients, $\dot{\mathbf{r}}_i$ and $\boldsymbol{\omega}_i$ are the translational and angular velocities of particle $i$, and m and $I$ are the mass and moment of inertia. The stochastic translational and rotational forces $\boldsymbol{\xi}^T$ and $\boldsymbol{\xi}^R$ satisfy the fluctuation dissipation theorem:

$$\langle \boldsymbol{\xi}_i^T(t) \rangle = \mathbf{0} \tag{S29}$$

$$\langle \boldsymbol{\xi}_i^R(t) \rangle = \mathbf{0} \tag{S30}$$

$$\langle \boldsymbol{\xi}_i^T(t)\boldsymbol{\xi}_j^T(t') \rangle = 2\gamma^T k_B T \delta_{ij} \mathbf{I} \delta(t-t') \tag{S31}$$

$$\langle \boldsymbol{\xi}_i^R(t)\boldsymbol{\xi}_j^R(t') \rangle = 2\gamma^R k_B T \delta_{ij} \mathbf{I} \delta(t-t') \tag{S32}$$

where $\delta_{ij}$ is the Kronecker delta, $\delta(t-t')$ is the Dirac delta function, and **I** is the identity tensor.

The particles' dipole magnitude is scaled to match that of experimental observation for 20 nm super-paramagnetic iron oxide nanoparticles with saturated magnetic dipoles [6]. The dimensionless dipole strength [7] is defined as

$$\Gamma = \frac{m^2}{2\pi\sigma^3 k_B T} \tag{S33}$$

corresponding to the ratio of interaction energy of two perfectly aligned adjacent dipoles to the thermal energy $k_B T$. For all results shown here we use a dimensionless strength of $\Gamma = 2.84$ as determined from experimental measurements of $m$. The surface area fraction of particles is held constant at $\eta = N r_c^2 / R^2 = 0.8$. All simulations are conducted with a minimum of 6532 particles in LAMMPS [8].

After an initial equilibration run, a weak uniform external magnetic field is applied uniaxially in the z-direction to align particles. The resulting dipolar orientation is shown in Fig. 3a of the main text. The results indicate a strong spatial dependence of the dipolar orientation, with diminishing alignment with increasing distance from the equator.

To further understand the magnetic environment beyond the droplet surface, we compute the contribution to the magnetic field energy induced by the surface particles in both the interior and exterior of the droplet. We do this by placing ideal dipoles pointing in the field direction throughout the droplet interior, exterior and surface and measuring the net dipolar energy felt by a test particle through interactions with surface particles. The resulting energies are then converted to the z-component of the induced field by dividing by the square of the dipolar magnitude. Due to the spherical symmetry of the system, the induced field can be projected on to the x-z plane as represented in Fig. 3b (main text). The shell of the droplet displays a quadrupolar interaction field while the magnetic field of the exterior of the droplet exhibits another quadrupolar field that is reversed in polarity relative to the shell's field.

The results from continuum analysis in Fig. 3c (main text) show chemical potential contours similar to the field gradients computed from simulations, shown in Fig. 3b (main text). Importantly, the chemical potential is significantly lower within the shell in comparison to the droplet interior, suggesting a strong magnetic driving force for interfacial assembly in the presence of a uniform magnetic field. This result, in conjunction with the experimentally observed reduction in the interfacial tension with applied field, provides compelling evidence for increased NPS interfacial assembly in the presence of a magnetic field. We note that the chemical potential varies spatially within the shell, with high chemical potential at the poles and a lower chemical potential near the equator. Such a spatially inhomogeneous chemical potential may drive density variations of the NPS at the interface. These induced density variations may be important towards a complete understanding of explosive emulsification and is the subject of ongoing investigation.

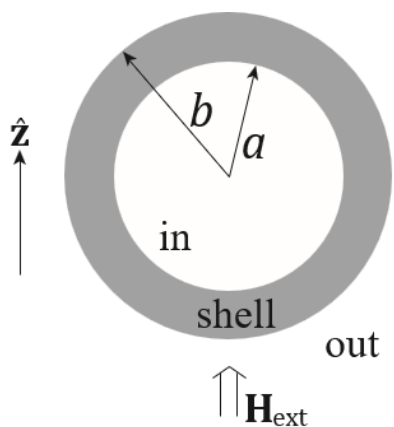

**Supplementary Fig. 1. Spherical shell geometry for continuum linear response calculation.**

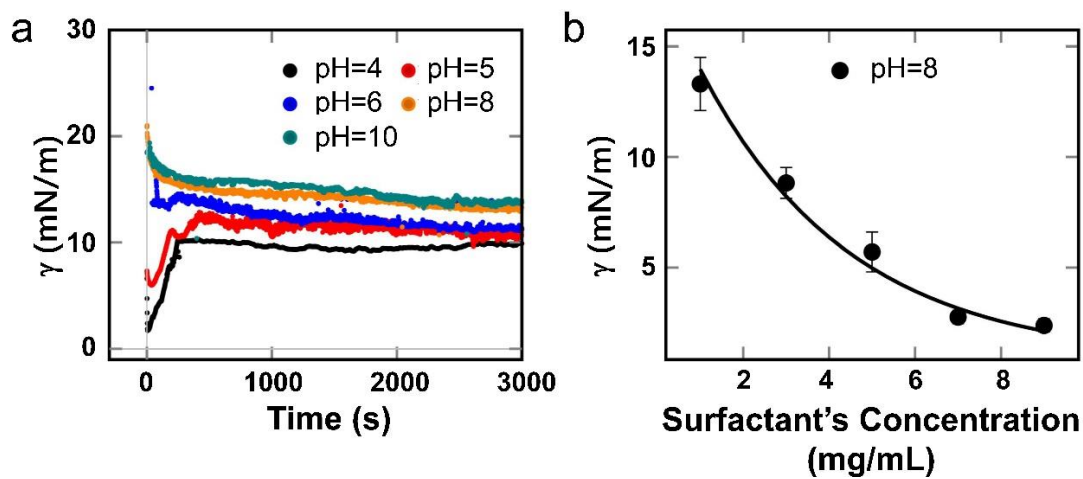

**Supplementary Fig. 2. The measurement of the interfacial tension.**

**a,** The time evolution of interfacial tension (γ) between the aqueous and toluene phase as a function of pH. **b,** The equilibrium interfacial tension between the aqueous and toluene phase as a function of surfactant's concentration. The concentrations of $Fe_3O_4$ nanoparticles and PS-triNH$_2$ both are 1 mg/mL.

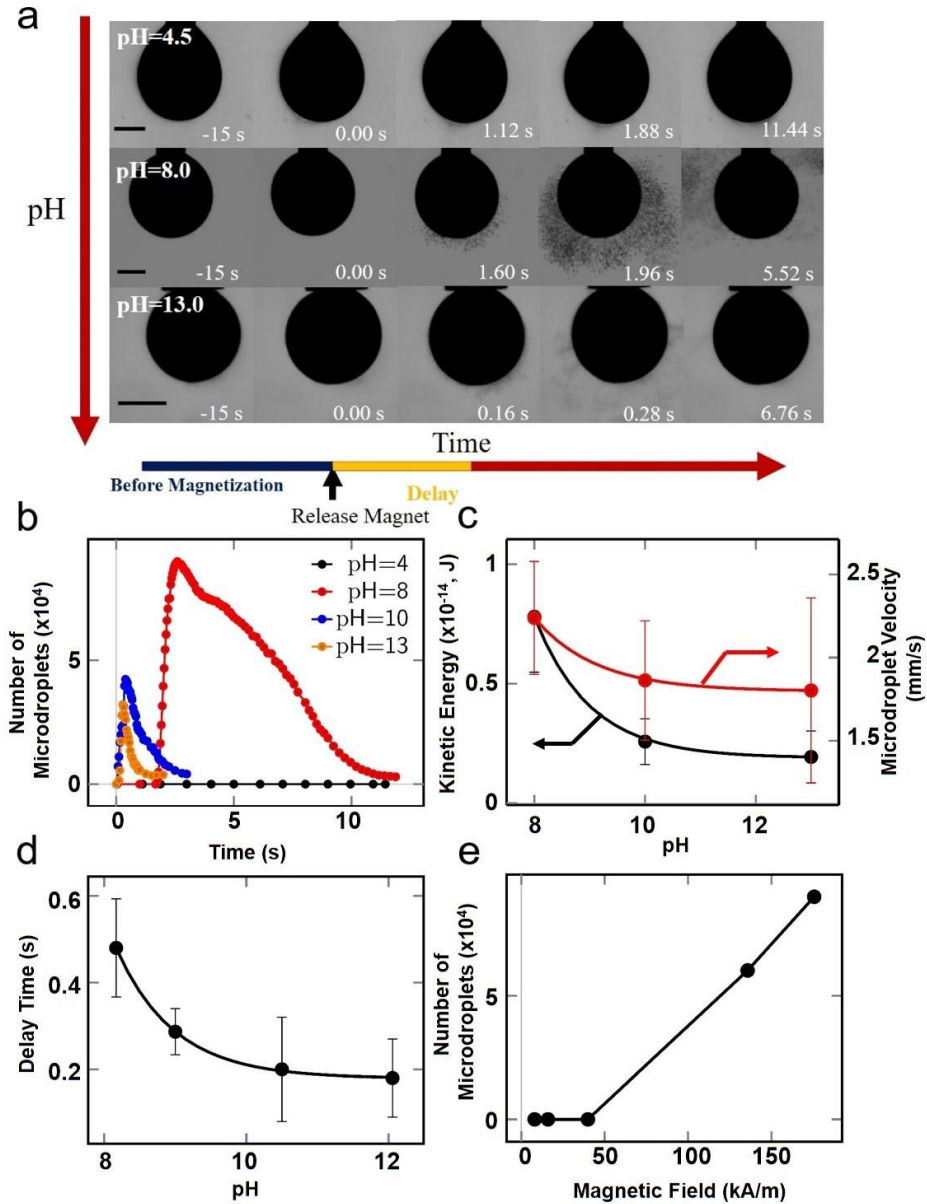

**Supplementary Fig. 3. Explosive emulsification as a function of pH.**

**a,** The optical images of the aqueous droplets during and after magnetization at different pH. **b,** The temporal changes of the ejected microdroplets' number, $N$, at different pH. **c,** Ejection velocity, $v$, and total kinetic energy, $E$, of the ejected microdroplets as a function of pH. The total kinetic energy is calculated using the formula: $E = \frac{1}{2} m_{microdroplet} N v^2$, where $m_{microdroplet}$ is the mass of per microdroplet. **d,** Effect of pH on the length of delay. **e,** The ejected microdroplets' number triggered at different magnetic field intensity and the pH of aqueous droplet is 8. The concentrations of $Fe_3O_4$ nanoparticles and PS-triNH$_2$ both are 1 mg/mL. Scale bar is 500 μm.

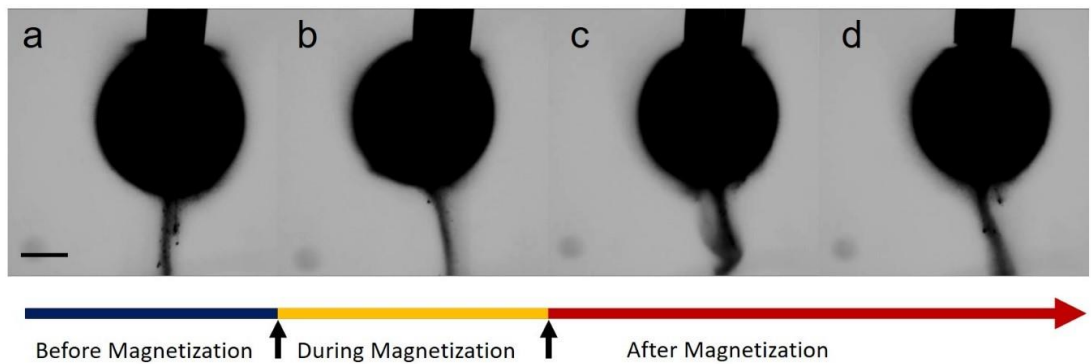

**Supplementary Fig. 4. The time serials optical images of the pendant aqueous droplet before, during and after magnetization.**

The concentrations of $Fe_3O_4$ nanoparticles and PS-triNH$_2$ are 1 mg/mL and 9 mg/mL, respectively. The pH of aqueous phase is 7. Scale bar: 500 μm.

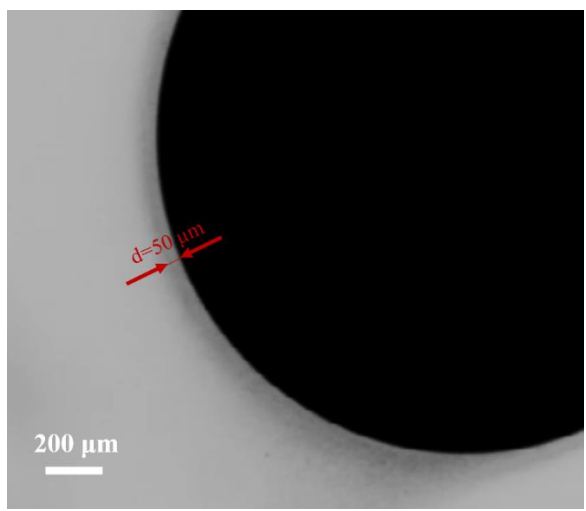

**Supplementary Fig. 5. The optical microscopy image of the aqueous droplet/oil interface.**

The concentrations of $Fe_3O_4$ nanoparticles and PS-triNH$_2$ both are 1 mg/mL. The pH of aqueous phase is 5.

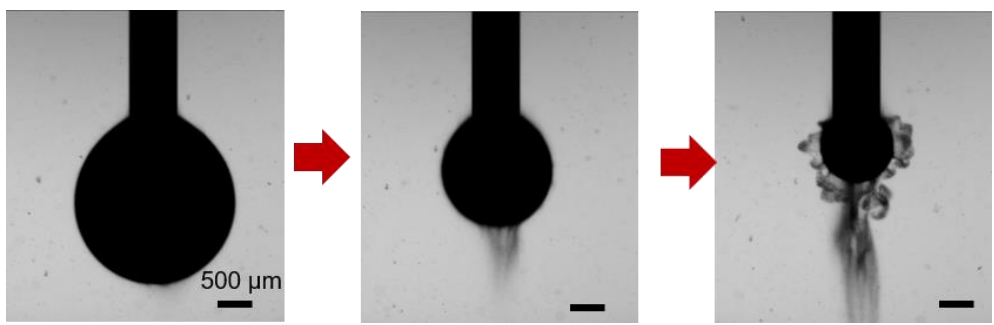

**Supplementary Fig. 6 The time series snapshots of the pendant aqueous droplet immersed in the toluene phase during a reduction of the droplet volume.**

From the left to right, the time series is selected as t= 0 s, 5 s and 15 s. The concentrations of $Fe_3O_4$ nanoparticles and PS-triNH$_2$ both are 1 mg/mL. The pH of aqueous phase is 8.

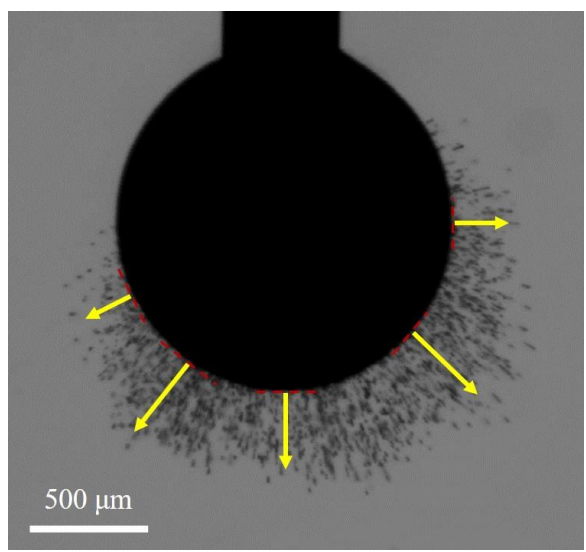

**Supplementary Fig. 7. The optical graph of an aqueous droplet occurring explosive emulsification.**

The red dash lines represent the tangent lines at the surface, and the yellow arrows represent the ejection direction of the emulsions.

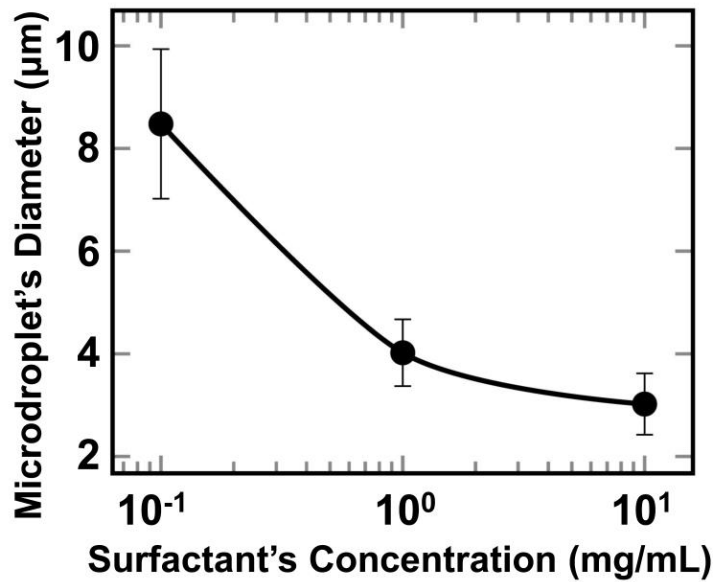

**Supplementary Fig. 8. The size of the microdroplet varying with the concentration of the surfactant in oil phase.**

The concentrations of $Fe_3O_4$ nanoparticles are 1 mg/mL, and the pH of the aqueous phase is 5.

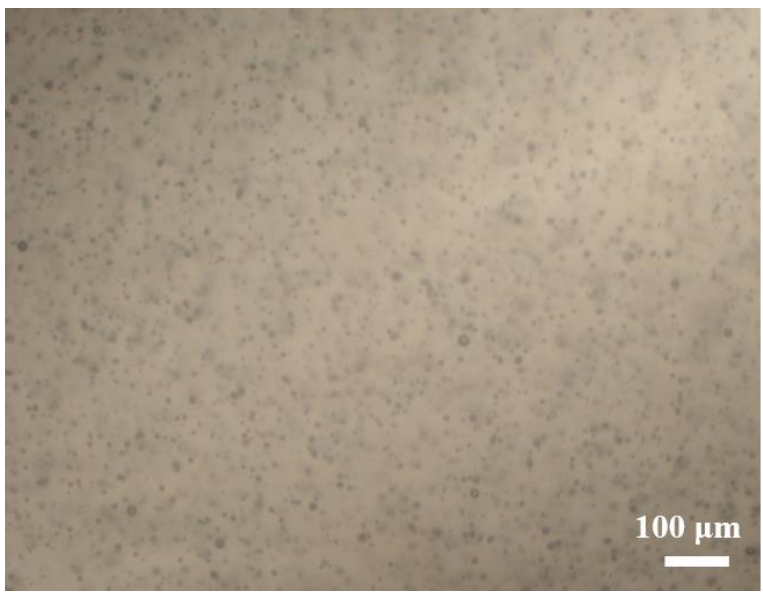

**Supplementary Fig. 9. Micrograph of the emulsions stably dispersed in the toluene.**

The concentrations of $Fe_3O_4$ nanoparticles and PS-triNH$_2$ both are 1 mg/mL.

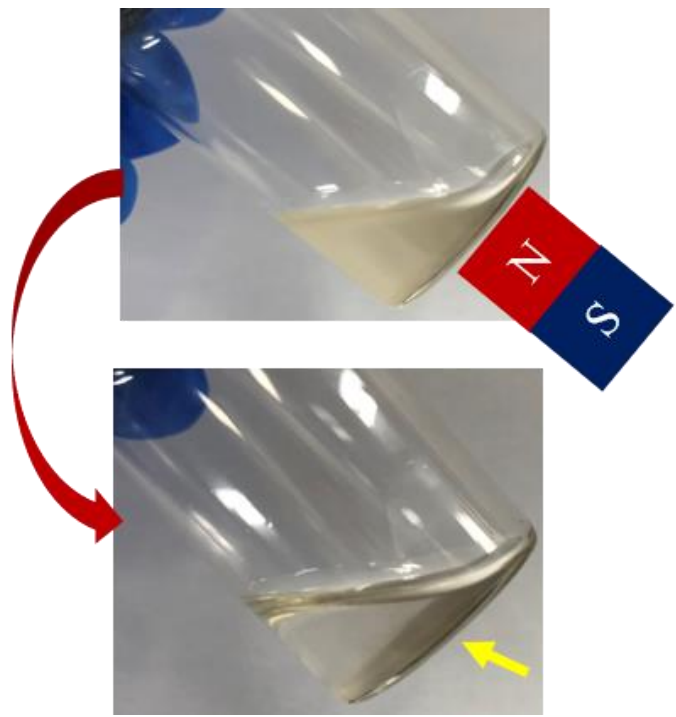

**Supplementary Fig. 10. The photographs of the microdroplet dispersions before and after magnetization by the magnet below the vial.**
The yellow arrow indicates the microdroplet are attracted by the magnet to the bottom.

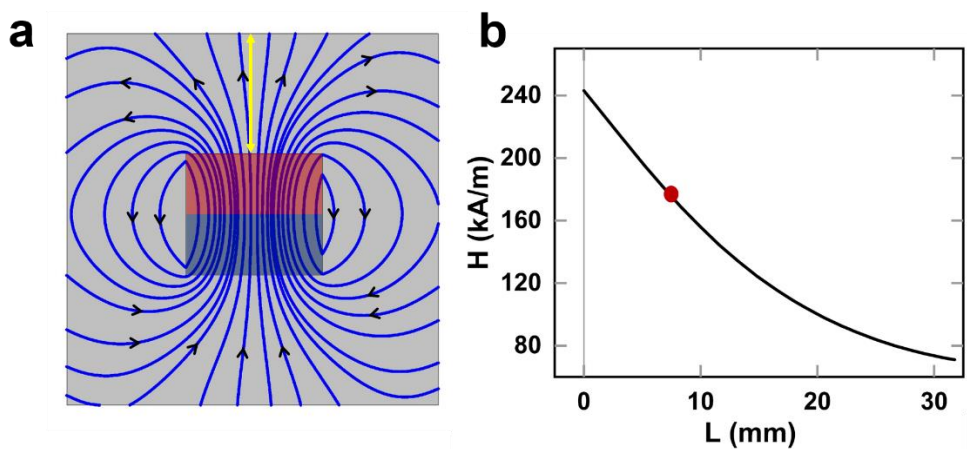

**Supplementary Fig. 11 2D simulation results of the magnetic field surrounding the NdFeB magnet (36 mm*32 mm*36 mm).**
**a,** The magnetic field lines surrounding the magnet; **b**, the relationship between the magnetic field intensity (H) and the distance away from the surface of the magnet (L) (yellow line in **a**). The red dot represents the measured distance between the pendant droplet and magnet (~7 mm), where the magnetic field intensity is up to 175 kA/m. (COMSOL Multiphysics).

**Supplementary Video 1.**

Explosive emulsification triggered by the external magnetic field. The concentrations of $Fe_3O_4$ nanoparticles and PS-triNH$_2$ are 1 mg/mL and 1 mg/mL, respectively. The pH of aqueous phase is 8.

**Supplementary Video 2.**

Spontaneous emulsification. The concentrations of $Fe_3O_4$ nanoparticles and PS-triNH$_2$ are 1 mg/mL and 7 mg/mL, respectively. The pH of aqueous phase is 7.

**Supplementary Video 3.**

The movement of microdroplets on a magnet stirrer. The concentrations of $Fe_3O_4$ nanoparticles and PS-triNH$_2$ are 1 mg/mL and 5 mg/mL, respectively. The pH of aqueous phase is 7. Rotating microdroplets indicated by arrows.